\documentclass[journal,onecolumn,12pt,twoside]{IEEEtranTCOM}
\usepackage{amssymb,amsmath,epsfig}
\usepackage{graphicx,times,latexsym}
\usepackage{psfrag,cite}
\usepackage{setspace}
\usepackage{algorithmic, algorithm}
\DeclareMathOperator*{\argmax}{arg\,max}
\DeclareMathOperator*{\argmin}{arg\,min}
\onecolumn
\begin{document}

\title{Performance Analysis of Network Coded Systems Under Quasi-static Rayleigh Fading Channels}

\author{Tu\u{g}can~Akta\c{s},~\IEEEmembership{Student~Member,~IEEE,}
        Ali~\"{O}zg\"{u}r~Y\i lmaz,~\IEEEmembership{Member,~IEEE,}
        and~Emre~Akta\c{s},~\IEEEmembership{Member,~IEEE}
\thanks{Tu\u{g}can~Akta\c{s} and {A.~\"{O}zg\"{u}r~Y\i lmaz} are with the Dept. of Electrical and Electronics Engineering, Middle East Technical University (taktas@metu.edu.tr, aoyilmaz@metu.edu.tr) and {Emre~Akta\c{s}} is with the Dept. of Electrical and Electronics Engineering, Hacettepe University (aktas@ee.hacettepe.edu.tr), Ankara, Turkey. This work was presented in part at the 2013 IEEE International Symposium on Information Theory (ISIT 2013), Istanbul, Turkey.}}
\markboth{IEEE Transactions on Communications}%
{Submitted paper}
\maketitle

\begin{abstract}
In the area of basic and network coded cooperative communication, the expected end-to-end bit error rate (BER) values are frequently required to compare the proposed coding, relaying, and decoding techniques. Instead of obtaining these values via time consuming Monte Carlo simulations, deriving closed form expressions using approximations is crucial. In this work, the ultimate goal is to derive an approximate average BER expression for a network coded system. While reaching this goal, we firstly consider the cooperative systems' instantaneous BER values that are commonly composed of Q-functions of more than one variables. For these Q-functions, we investigate the convergence characteristics of the \textit{sampling property} and generalize this property to arbitrary functions of multiple variables. Second, we adapt the equivalent channel approach to the network coded scenario for the ease of analysis and propose a network decoder with reduced complexity. Finally, by combining these techniques, we show that the obtained closed form expressions well agree with simulation results in a wide SNR range.

\end{abstract}

\IEEEpeerreviewmaketitle

\IEEEpubidadjcol

\section{INTRODUCTION}

In the area of wireless communications, especially with low-power nodes that are not capable of accommodating multiple antennas, making use of cooperation between nodes is a preferred method for creating spatial diversity resulting from the broadcast nature of the communication medium. Through cooperation between nodes, the overall reliability and the throughput of the system can be improved by mitigating the deteriorating effects of fading. The basic cooperation scenario assumes a dedicated relay node assisting the communication between a source and a destination node~\cite{Erkip03}. Out of various types of relaying, Demodulate-and-Forward (DMF) is preferable with low-power relay nodes thanks to its low computational complexity. This low complexity takes its roots from the fact that a DMF-type relay simply hard-detects the incoming symbols and forwards these detection results to the destination node without any channel decoding operation~\cite{Laneman04}. On the other hand, in order to avoid the error propagation problem due to DMF method, one has to take the possible relay errors into consideration at the destination node. In this respect, one may prefer to use the reliability information of the source-relay channel and make a maximum likelihood (ML) detection at the destination side~\cite{Laneman06} or adhere to the equivalent channel approach in conjunction with the cooperative maximal-ratio combining (C-MRC) method which reaches a decision based on a weighted sum of observations again by utilizing the source-relay channel reliability information~\cite{Wang07}. Both of these techniques are shown to achieve full-diversity order with DMF-type relaying.

One of the first studies on network coding considered error-free wired links and proved that coding at the intermediate (relay) nodes may improve the information flow rate in the network~\cite{Ahlswede00}. Later many studies investigated the properties and limits of network coding strategies for both wired and wireless operations~\cite{Renzo10}. In network coded wireless communications, which can be seen as an extension of the basic cooperative communication scenario, the throughput in the whole network is shown to be improved by combining data packets of many sources in a single packet via Galois Field (GF) operations at the intermediate node~\cite{Medard08}. In this study, we aim to devise analysis methods for the BER performance of wireless network coding scenarios with DMF-type intermediate nodes under quasi-static Rayleigh fading channels by starting our analysis from the basic relayed communication model. 

Regarding performance of the equivalent channel approximation together with the C-MRC method at the destination side, the diversity order analysis for the basic relayed communication scenario under Rayleigh fading is done in~\cite{Wang07} and~\cite{Kim2011}, where the authors reach the result that the related system achieves a diversity order of $2$. This result is obtained by applying a number of approximation techniques on the end-to-end instantaneous and average BER expressions and the final average BER expressions are very loose in general. Recently in~\cite{Uysal2013}, authors use C-MRC for the analysis of single-relay network coded communication. Here, in this work, one of our goals is deriving a closed-form approximate average BER expression for the basic relayed communication scenario, which is tight in the mid-to-high SNR region and that gives us the coding gain term in addition to the previously found diversity order term. In obtaining this closed-form expression, we make use of the sampling property of Q-function introduced in~\cite{Jang2011,Jang2011TCOM}, where the expectation integrals required for investigating the average BER performance of cooperative systems are handled through an approximation on the Q-function appearing in the instantaneous BER functions. We generalize the sampling property to more general arguments of Q-function with possibly more than one variables. Such a general form of sampling property is avoided in~\cite{Jang2011,Jang2011TCOM} by expressing any function of more than one variables as sum of single-variable functions through approximations which yield coding gain offsets in the final expression. Moreover, we try to characterize the low-SNR region approximation problem with the sampling property, which is also pointed in~\cite{Jang2012}. Different than~\cite{Jang2012}, we analyze the \textit{convergence rates} for the constituent functions of the integrand function in order to distinguish a threshold value above which the Q-function related part can safely be approximated by a Dirac-delta generalized function. Finally, we adopt the equivalent channel approach and the C-MRC technique to the network coded scenario with multiple intermediate nodes introduced in~\cite{Aktas2013} and analyze the average BER for a sample network again by using the generalized sampling property for the Q-function that we propose. Both for the basic cooperative system and the sample network coded cooperative system, we compare the closed-form expressions we derived with the simulated BER curves and observe a very good agreement between them.

The rest of the paper is organized as follows. Section~\ref{sec:model} introduces the basic cooperative communication model, the equivalent channel approach,the C-MRC technique and the related end-to-end instantaneous BER expression. In Section~\ref{sec:Qfunc}, we start with a brief introduction to the sampling property for expectation integrals and continue with an investigation on the SNR region for which this property is valid. Furthermore, we generalize the sampling property of the Q-function to more than one dimensions and obtain a closed-form expression for the basic cooperative model. Next, in Section~\ref{sec:nwc}, the analysis and the detection methods detailed in previous sections are generalized to the network coded communication systems in order to reach the corresponding average BER expressions in closed form. The agreement of the derived expressions with the simulations is presented in Section~\ref{sec:sim_results}. Section~\ref{sec:concl} draws the conclusions and pose some possible paths for future work.

\section{Canonical Cooperative Communication Model and Instantaneous End-to-End BER Expression}
\label{sec:model}

\begin{figure}[htbp]
   \centering
   \includegraphics[width=0.46\textwidth]{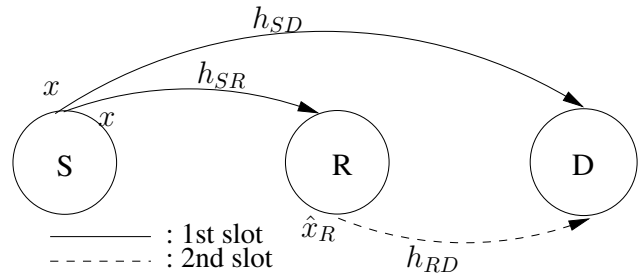}
   \caption{Canonical cooperative communication system}
   \label{fig:basic_coop}
\end{figure}

The cooperative communication system which is composed of a source, a destination and a relay node assisting these nodes is referred as the canonical cooperative communication system. This system and related fading coefficients assigned to the links between the source node $S$, the relay node $R$ and the destination node $D$ are presented in Fig.~\ref{fig:basic_coop}. The channel fading coefficients $h_{SR}$, $h_{RD}$, and $h_{SD}$ are assumed to be independent and to follow zero mean circularly symmetric complex Gaussian probability distributions such that $h_{ij}\sim CN(0,\sigma_{ij}^2)$, where $(ij) \in \{SR, RD, SD\}$, nodes $S$ and $R$ access the channel in an orthogonal fashion according to a time-division method without loss of generality so that in the first time slot $S$ transmits the data symbol $x$ and nodes $R$ and $D$ have respective observations
\vskip-\parskip 
\begin{equation}
\label{eqn:obs}
y_{SR}=h_{SR} x+n_{SR}\ \text{ and } \  
y_{SD}=h_{SD} x+n_{SD},
\end{equation}
where $n_{SR}$ and $n_{SD}$ denote the independent white complex Gaussian noise terms at $R$ and $D$ with identical distribution $CN(0,N_0)$. 
\vskip-\parskip

As in~\cite{Wang07}, Binary Phase Shift Keying (BPSK) is assumed for transmissions for analytical tractability with $x\in \{ +\sqrt{E},-\sqrt{E}\}$, where $E$ is the average bit energy. Next, we define $\bar{\gamma}=\frac{E}{N_0}$ so that the instantaneous SNR values for $S$-$R$ and $S$-$D$ links are $\gamma_{SR}=\bar{\gamma} \vert h_{SR}\vert ^2$ and $\gamma_{SD}=\bar{\gamma} \vert h_{SD}\vert ^2$ respectively. These instantaneous values are exponentially distributed with respective expectations $\bar{\gamma} \sigma_{SR}^2$ and $\bar{\gamma} \sigma_{SD}^2$. When $R$ operates as a DMF-type relay, it first detects
\vskip-\parskip 
\begin{equation}
\label{eqn:relay_det}
\hat{x}_R=\argmin_{x\in \left\lbrace +\sqrt{E},-\sqrt{E} \right\rbrace } \vert y_{SR} - h_{SR}x\vert ^2
\end{equation}
and simply transmits $\hat{x}_R$, which is the optimal hard detection result for $x$. The observation of $D$ after the second time slot is then $y_{RD}=h_{RD} \hat{x}_R+n_{RD}$, where $n_{RD}\sim CN(0,N_0)$ and the corresponding instantaneous SNR $\gamma_{RD}=\bar{\gamma} \vert h_{RD}\vert ^2$ is also exponentially distributed.

It is given in~\cite{Wang07} that the C-MRC method at node $D$ has less computational complexity and is analytically more tractable with respect to the optimal detection rule. Moreover, through simulations it is shown to perform very close to the optimal rule. By using C-MRC, the detection is carried out on the weighted sum of two observation signals (one directly from the source, the other over the relay). The weighting is done in accordance with the reliability of the two receptions
\vskip-\parskip
\begin{equation}
\label{eqn:detD}
\hat{x}_D=\argmin_{x\in \{ +\sqrt{P},-\sqrt{P}\} } \vert w_1 y_{SD} + w_2 y_{RD} - (w_1 h_{SD} + w_2 h_{RD})x\vert ^2,
\end{equation}
where $w_1$ is the weight coefficient corresponding to the $S-D$ link and is equal to $h_{SD}^\ast$ as in the well-known MRC method without relaying. On the other hand, for observation $y_{RD}$, which corresponds to the relayed communication over the links $S-R$ and $R-D$, the coefficient $w_2$ should be re-defined to reflect the possible error propagation on these two hops. In~\cite{Wang07}, the authors propose a single equivalent channel for representing these hops and approximate the equivalent instantaneous SNR of this channel with
\vskip-\parskip
\begin{equation}
\label{eqn:gamma_eq}
\gamma_{eq}\triangleq \left\lbrace Q^{-1}\left( \left[1-P_{SR}^b\right] P_{RD}^b + \left[1-P_{RD}^b\right] P_{SR}^b\right)\right\rbrace^2 /2, 
\end{equation}
where $Q(x)=\int_{x}^{\infty}\frac{1}{\sqrt{2\pi}}\exp(-z^2/2)\text{d}z$, $Q^{-1}$ is the inverse Q-function, $P_{SR}^b=Q(\sqrt{2\gamma_{SR}})$ and $P_{RD}^b=Q(\sqrt{2\gamma_{RD}})$. Accordingly we define $w_2 =\frac{\gamma_{eq}}{\gamma_{RD}}h_{RD}^\ast$ and the instantaneous end-to-end BER expression for detection in (\ref{eqn:detD}) is found as~\cite{Wang07}
\vskip-\parskip
\begin{align}
\label{eqn:P_b}
P^b =& \left( 1 - Q(\sqrt{2\gamma_{SR}})\right)Q\left[ \frac{\sqrt{2}(\gamma_{SD}+\gamma_{eq})}{\sqrt{\gamma_{SD}+\gamma_{eq}^2/\gamma_{RD}}}\right]
+ Q(\sqrt{2\gamma_{SR}})Q\left[ \frac{\sqrt{2}(\gamma_{SD}-\gamma_{eq})}{\sqrt{\gamma_{SD}+\gamma_{eq}^2/\gamma_{RD}}}\right].
\end{align}
In order to obtain the expected end-to-end BER, one has to evaluate a triple integral for the instantaneous BER function in (\ref{eqn:P_b}) over the related distributions of the random variables $\gamma_{SR}$, $\gamma_{SD}$, $\gamma_{RD}$ (here, $\gamma_{eq}$ is a function of $\gamma_{SR}$ and $\gamma_{RD}$). However, accomplishing this analytically is hard and also the alternative method of resorting to Monte Carlo simulations is time consuming in general. In~\cite{Wang07}, only the diversity order of the average BER expression is obtained following a series of upper bounding techniques and the result is $2$ since two transmissions are made on independent paths in the network and $D$ accounts for possible relaying errors through C-MRC. In Section \ref{sec:Qfunc_relay}, we present novel closed form expressions for the average BER of this system in which the coding and the diversity gains are identified separately.

\section{Sampling Property of the Q-Function for Generalized Expressions}
\label{sec:Qfunc}
In this section, firstly, we propose a simple method in order to improve the sampling property of the Q-function that is essentially presented in~\cite{Jang2011TCOM}. We continue with demonstrating the discrepancy with this sampling property especially in the low-SNR region. By analysis over the constituent functions involved in the expectation integral, we remedy this deficiency and generalize the sampling property for low-SNR region as well. Finally, the sampling property is further generalized to expectation integrals whose integrand functions involve more than one variables.

\subsection{Basic Problem and its Solution}
\label{sec:Qfunc_single}
Assume that the following expectation integral of an instantaneous probability of error function $Q(\sqrt{X})$ is to be evaluated for a random variable $X$ with probability density function (pdf) $f_X(x)$~\cite{Jang2011TCOM}:
\vskip-\parskip
\begin{equation}
\label{eqn:Qsqrtx}
I_0 = E_X\left\lbrace Q(\sqrt{X}) \right\rbrace = \int_{0}^{\infty}Q(\sqrt{x})f_X(x)\text{d}x.
\end{equation}
After the change of variables operation $x\rightarrow t^N$, we have the integrand $Q(\sqrt{t^N})Nt^{N-1}f_X(t^N)$. Here we define the following constituent functions of the integrand.
\vskip-\parskip
\begin{equation}
\label{eqn:prod_defn}
q(t)\triangleq Q(\sqrt{t^N}), \;c(t) \triangleq Nt^{N-1},\; f(t)\triangleq f_X(t^N).
\end{equation}
In~\cite{Jang2011TCOM}, $h(t^N)\triangleq q(t)c(t)$ is defined and claimed to be a unimodal function of $t$ with a critical point satisfying $t^N_{\ast}=2$ as ${N\rightarrow\infty}$. However, that analysis does not show that $h(t^N)$ assumes the value of zero at all other points. On the contrary, when any other finite $t^N$ value is inserted in Eqn. (52) of~\cite{Jang2011TCOM} it is easy to show that  $h(t^N)$ assumes infinity. Moreover, the critical point $t^N_{\ast}=2$ is obtained after applying an upper bound on the Q-function. Here, we will take a different approach without any approximations to show that $h(t^N)$ indeed assumes an infinite value only around the point $t=1$ and converges to $0$ everywhere else. In addition, we suggest an alternative way to calculate  $t^N_{\ast}$ so that the low-SNR agreement with the simulation results is enhanced. We then propose a piecewise sampling method on constituent functions to further improve our technique.

Let us start with the analysis of the integrand $I(t)\triangleq q(t) c(t) f(t)$ for three distinct regions of $t$:
\vskip-\parskip
\begin{equation}
\label{eqn:Qsqrtx_3region}
I_0 = \int_{0}^{1^-}I(t)\text{d}t + \int_{1^-}^{1^+}I(t)\text{d}t + \int_{1^+}^{\infty}I(t)\text{d}t.
\end{equation}
It should be emphasized that the value $I_0$ in (\ref{eqn:Qsqrtx_3region}) is independent of $N$ and hence we may investigate the behaviour of $I(t)$ asymptotically (as ${N\rightarrow\infty}$). In this work, we take $f(t)=\frac{1}{SNR}e^{-\frac{t^N}{SNR}}$ (Rayleigh fading assumption) although the following steps can be generalized to other pdfs. Initially, for the $0<t<1$ region, we have 
\vskip-\parskip
\begin{align}
\label{eqn:t_less_1}
\lim_{N\rightarrow\infty}I(t) \bigg\vert_{0<t<1}&= \left(\lim_{N\rightarrow\infty}q(t)\right) \left(\lim_{N\rightarrow\infty}c(t)\right) \left(\lim_{N\rightarrow\infty}f(t)\right)\nonumber\\
&= \left(\frac{1}{2}\right) \left(\lim_{N\rightarrow\infty}\frac{1}{\ln{t} (-t^{1-N})}\right) \left(\frac{1}{SNR}\right) \;\nonumber\\
&= 0,
\end{align}
where the L'H\^{o}pital rule is applied for $\lim_{N\rightarrow\infty}c(t)$ term and the product law for limits is utilized. As a result, the first integral in (\ref{eqn:Qsqrtx_3region}) evaluates to $0$ asymptotically. Next, for the $t>1$ region, we have 
\vskip-\parskip
\begin{align}
\label{eqn:t_larger_1}
\lim_{N\rightarrow\infty}I(t) \bigg\vert_{t>1} &= \left(\lim_{N\rightarrow\infty}q(t)\right) \left(\lim_{N\rightarrow\infty}c(t)f(t)\right)\nonumber\\
&= \left(0\right) \left(\lim_{N\rightarrow\infty}\frac{N t^{N-1}}{\exp\left( \frac{t^N}{SNR}\right)}\right) \left(\frac{1}{SNR}\right)\nonumber\\
&= \left(0\right) \left(\lim_{N\rightarrow\infty}\frac{t^{N-1} + N (\ln{t}) t^{N-1}}{\frac{\ln{t}}{SNR}t^N\exp\left( \frac{t^N}{SNR}\right)}\right)\nonumber\\
&= \left(0\right) \left(\lim_{N\rightarrow\infty}\frac{\ln{t}}{\left(\frac{\ln{t}}{SNR}\right)^2 t^N\exp\left( \frac{t^N}{SNR}\right)}\right) \nonumber\\
&= 0
\end{align}
following application of the L'H\^{o}pital rule twice. Hence, in the asymptotic sense, the third integral in (\ref{eqn:Qsqrtx_3region}) does not contribute to the result either. Therefore, we obtain
\vskip-\parskip
\begin{equation}
\label{eqn:Qsqrtx_1region}
I_0 = \int_{1^-}^{1^+}\lim_{N\rightarrow\infty} I(t)\text{d}t,
\end{equation}
which shows us that the integrand $I(t)$ may be well-approximated by a Dirac delta generalized function at $t=1$ for ${N\rightarrow\infty}$. Moreover, it can be shown that this Dirac delta approximation also holds for the functions $h(t^N)\triangleq q(t)c(t)$ and $g(t^N)\triangleq f(t)c(t)$ by following the arguments utilized in reaching (\ref{eqn:t_less_1}) and (\ref{eqn:t_larger_1}). Hence it is important to characterize the functions $h(t^N)$ and $g(t^N)$ in the regions $t<1$ and $t>1$ in order to select the one with faster decay rate. Next, we are going to identify an SNR value above which $h(t^N)$ can be safely approximated by a Dirac delta in Section~\ref{sec:comp_conv}.
\subsection{Rates of Convergence for Constituent Functions}
\label{sec:comp_conv}

In order to approximate the basic integral of (\ref{eqn:Qsqrtx}) by using the sampling property, we need to obtain two essential parameters of the Dirac delta generalized function, which is an approximation to one of the constituent functions $h(t^N)$ or $g(t^N)$. The first parameter is the location of Dirac delta, $t^N_{\ast}$. The asymptotic (as $N\rightarrow \infty$) critical point of the function of interest yields this parameter and the critical point can be either found analytically or by solving a simple unconstrained optimization problem. The other parameter is the weight of the function $c$ and is analytically obtained for both of the constituent functions by integrating the related function from $0$ to $\infty$. However, another important issue is to pick the sampling function as one of $h(t^N)$ or $g(t^N)$ according to their asymptotic convergence rates to the Dirac delta.

We start with comparing the convergence rates of $h(t^N)$ and $g(t^N)$ for $0<t<1$ and $t>1$, separately. Firstly, consider the region $0<t<1$. For this region, both $q(t)$ and $f(t)$ converge to nonzero constants in the limit. On the other hand, the function $c(t)=Nt^{N-1}$ converges to $0$, which means that for large $N$, $h(t^N)$ and $g(t^N)$ converge to $0$ with the same rate. Hence we focus on the other region: $t>1$. In this region we compare $h(t^N)$ and $g(t^N)$ starting with the Chernoff upper bound on the Q-function:
\vskip-\parskip
\begin{equation}
\label{eqn:q_chernoff}
q(t)\bigg\vert_{t^N=x} = Q(\sqrt{x}) \leq \frac{1}{2} \exp\left( -\frac{x}{2}\right).
\end{equation}
In addition, for $t>1$ and $SNR \geq 2$ it is easily shown that 
\begin{equation}
\label{eqn:exp_ineq}
f(t)\bigg\vert_{t^N=x} = \frac{\exp\left(\frac{-x}{SNR}\right)}{SNR} \geq \frac{1}{2} \exp\left( -\frac{x}{2}\right).
\end{equation} 
Combining (\ref{eqn:q_chernoff}) and (\ref{eqn:exp_ineq}) for $t>1$ and $SNR \geq 2$ (in dB scale roughly for values larger than $3$ dB) we get
\begin{equation}
\label{eqn:q_exp_comp}
q(t)\bigg\vert_{t^N=x} \leq f(t)\bigg\vert_{t^N=x}.
\end{equation}
Using (\ref{eqn:q_exp_comp}) we reach the result that for $SNR>2$, $h(t^N)$ (including the Q-function) is better represented by a Dirac delta with respect to $g(t^N)$ (including the exponential pdf). Also, for $t>1$ and $SNR<1/3$ (roughly less than $-5$ dB), one can show that 
\begin{equation}
\label{eqn:q_exp_comp2}
q(t)\bigg\vert_{t^N=x} = Q(\sqrt{x}) \geq \frac{1}{\frac{1}{3}} \exp\left( -\frac{x}{\frac{1}{3}}\right) \geq \frac{\exp\left(\frac{-x}{SNR}\right)}{SNR}=f(t)\bigg\vert_{t^N=x}
\end{equation}
Consequently, for lower SNR values, $g(t^N)$ is more suitable for the sampling function definition. Firstly, the position of the Dirac delta that approximates $g(t^N)$ can be obtained by finding the critical point $t_g ^N$. We equate the first derivative of $g(t^N)$ with respect to $t$ to $0$:
\vskip-\parskip
\begin{equation}
\frac{\text{d}}{\text{d}t}g(t^N)\bigg\vert_{t^N=t_g ^N}=0,
\end{equation}
whose solution is 
\begin{equation}
\label{eqn:exp_critical}
t_g ^N = \frac{N-1}{N}SNR.
\end{equation}
Eqn. (\ref{eqn:exp_critical}) gives us the asymptotic critical point $t_\ast ^N = \lim_{N\rightarrow\infty} t_g ^N=SNR$. Second, the weight of the corresponding Dirac delta is found as $1$ due to the normalization property of the pdf. Hence for $SNR<1/3$, we may use the approximation $I_0\approx \int_{0}^{\infty}Q(\sqrt{x})\delta(x-SNR)\text{d}x = Q\left( \sqrt{SNR}\right)$. For $SNR>2$, we write $I_0\approx \int_{0}^{\infty}c\delta(x-t_\ast ^N)f_X(x)\text{d}x = \frac{c}{SNR}\exp\left( -\frac{t_\ast ^N}{SNR}\right)$, where the impulse weight is found using the alternative definition of Q-function as $c=\int_{0}^{\infty}h(t^N)\text{d}t=\int_{0}^{\infty}Q(\sqrt{x})\text{d}x=\frac{1}{2}$. 
\begin{figure}[htbp]
   \centering
   \includegraphics[width=0.7\textwidth]{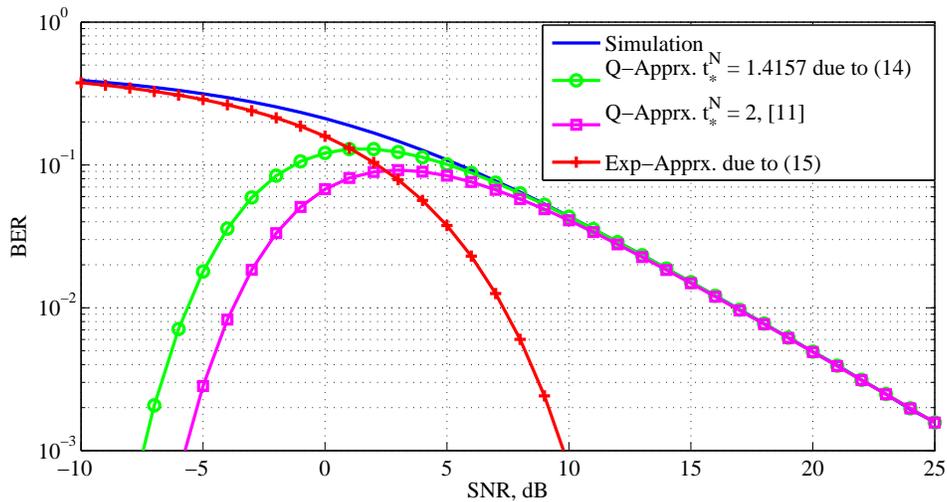}
   \caption{Approximating the integral $I_0$ using various methods}
   \label{fig:q_exp_comp}
\end{figure}

We propose a simple alternative to the method in~\cite{Jang2011TCOM} to evaluate the location of the impulse which approximates $h(t^N)$. For a sufficiently large value of $N$, we pose finding the critical point as an unconstrained optimization problem and employ numerical search to find the solution. In a narrow neighbourhood of $t=1$, we search for the critical point of $h(t^N)$, as an example by using the Optimization Toolbox function \textit{fminsearch()} of MATLAB. In MATLAB, we find $t_\ast ^N = 1.4157$ just after $4$ iteration steps, whereas in~\cite{Jang2011TCOM} $t_\ast ^N = 2$ was considered. The error rate curves from approximations as well as simulations are given in Fig.~\ref{fig:q_exp_comp}. According to Fig.~\ref{fig:q_exp_comp}, the methods approximating $h(t^N)$ as a Dirac delta (one proposed in~\cite{Jang2011TCOM} with square markers, and the one we propose with circular markers) are quite consistent for high SNR values. However, the method using unconstrained optimization for searching the impulse location is the better one with close approximation for $SNR>3$dB as detailed in equation (\ref{eqn:q_exp_comp}). For low SNR values, on the other hand, only the method selecting $g(t^N)$ as the sampling function (plus shaped markers) is close to the simulation result. This shows us that for a close approximation of the integral $I_0$ over the whole SNR region, we need to use a piecewise function. In the following sections, with integrands of more than one variables, we are going to use only the sampling property for the Q-function since we are mostly interested in the high SNR regime.

\subsection{Two-Variable Sampling Property}
\label{sec:Qfunc_double}
In this section we base our discussion on the following integral involving two variables in the integrand.
\vskip-\parskip
\begin{equation}
\label{eqn:Qsqrtx_y}
I_1 = \int_{0}^{\infty}\int_{0}^{\infty}Q(\sqrt{a_1x +a_2y})f_{X}(x)f_{Y}(y)\text{d}x\text{d}y, 
\end{equation}
where $a_1$ and $a_2$ are positive constants. The form of the expectation integral given in (\ref{eqn:Qsqrtx_y}) is frequently observed for receivers collecting observations on two independent channels and combining these received signals according to MRC operation. In the scenario of basic relayed communication, similar expectation integrals are also encountered~\cite{Jang2011,Jang2011TCOM,Jang2012}, however no higher dimensional generalization for sampling property has been made in the literature as far as we know. In~\cite{Jang2011}, a two-dimensional integrand is approximated by two single variable integrands resulting in coding gain offsets in the final expressions. 

\begin{figure}[htbp]
   \centering
   \includegraphics[width=0.7\textwidth]{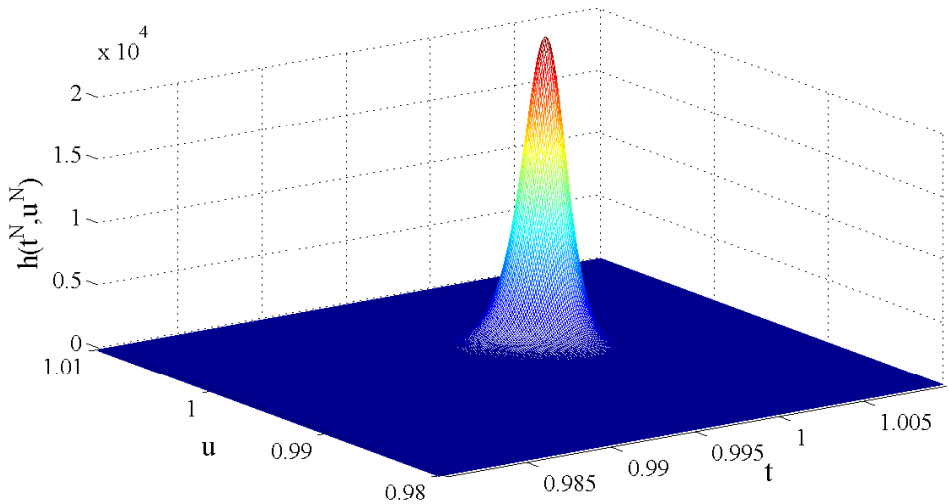}
   \caption{Function $h(t^N,u^N)$ for $N=1000$ and $a_1=a_2=2$}
   \label{fig:Qsqrtx_y}
\end{figure}

Similar to the single dimension analysis, we define $h(t^N,u^N)\triangleq Q(\sqrt{a_1t^N +a_2u^N})N^2t^{N-1}u^{N-1}$ based on the Q-function. We simply pick $h(t^N,u^N)$ as the sampling function, since it is shown to perform well in the high-SNR region in Section~\ref{sec:comp_conv}. Here, it is easy to generalize the asymptotic analysis for $h(t^N,u^N)$ with $N\rightarrow\infty$ to show that it is well-approximated by a two-dimensional Dirac delta at $(t,u)=(1,1)$. This is further exemplified in Fig.~\ref{fig:Qsqrtx_y} for $N=1000$ and $a_1=a_2=2$. 

Through the unconstrained optimization solution, the critical point of $h(t^N,u^N)$ is computed as $(t^N_{\ast},u^N_{\ast}) =(0.8197,0.8197)$ and the weight of the Dirac delta is analytically found as $c=\int_{0}^{\infty}\int_{0}^{\infty}Q(\sqrt{a_1x +a_2y})\text{d}x\text{d}y=\frac{3}{4a_1a_2}$. As an example for $a_1=a_2=2$, the approximation for $I_1$ is
\vskip-\parskip
\begin{equation}
\label{eqn:Qsqrtx_y2}
I_1 \approx \frac{3}{16SNR^2}\exp\left(-\frac{2(0.8197)}{SNR}\right).
\end{equation}
The result in (\ref{eqn:Qsqrtx_y2}) is in accordance with the average BER analysis result of the MRC technique applied on two parallel branches~\cite{Goldsmith} and is also very close to the simulation result for mid- to high SNR values as given in Fig.~\ref{fig:Qcombined}.

\subsection{Sampling in Single Dimension for Functions of Two Variables}
\label{sec:Qfunc_min}
Unfortunately, not every integrand function can be simply approximated with a two dimensional Dirac delta as in Section~\ref{sec:Qfunc_double}. As an example, the instantaneous BER function $Q(\sqrt{2\min\{x,y\}})$ can be investigated. This type of instantaneous BER function prevalently occurs in relayed communication system performance analysis, especially in the works that approximate the $S-R$ and $R-D$ links with a single link possessing the minimum one of the instantaneous SNR values of these links~\cite{Yuan2010, Yang2011, Wang07}. Here, it can be shown that the integrand $h(t^N,u^N)\triangleq Q(\sqrt{2\min\{t^N,u^N\}})N^2t^{N-1}u^{N-1}$ diverges also at points other than $(t,u)=(1,1)$. However, it is still possible to analyze this function using the fact that $Q(\sqrt{2\min\{x,y\}})\leq Q(\sqrt{2x}) + Q(\sqrt{2y})$ and the sampling property for single variable functions given in Section~\ref{sec:Qfunc_single}. Then we can reach the following approximation for this expectation integral, which is shown to perfectly fit the simulation result in Fig.~\ref{fig:Qcombined}.

\begin{align}
\label{eqn:Qsqrtminx_y}
I_2 =& \int_{0}^{\infty}\int_{0}^{\infty}Q(\sqrt{2\min\{x,y\}})f_{X}(x)f_{Y}(y)\text{d}x\text{d}y\nonumber\\
\approx& \frac{1}{2SNR}\exp \left( -\frac{0.7079}{SNR}\right).
\end{align}

\subsection{BER analysis for the Canonical Cooperative Model}
\label{sec:Qfunc_relay}
The end-to-end instantaneous BER function in (\ref{eqn:P_b}) can be written as the sum of two terms: $P^b=P_1+P_2$. Let us start with $P_1$:
\vskip-\parskip
\begin{equation}
P_1 = \left( 1 - Q(\sqrt{2\gamma_{SR}})\right)Q\left[ \frac{\sqrt{2}(\gamma_{SD}+\gamma_{eq})}{\sqrt{\gamma_{SD}+\gamma_{eq}^2/\gamma_{RD}}}\right], 
\end{equation}
which is a function of three variables, $\gamma_{SR}$, $\gamma_{RD}$, and $\gamma_{SD}$. Similar to the analysis in Section~\ref{sec:Qfunc_min}, function $P_1$ can not be approximated with a Dirac delta directly, due to the variable $\gamma_{eq}$ defined over the instantaneous SNR values of the two-hop link, $\gamma_{RD}$ and $\gamma_{SD}$. Therefore, we define two terms that are asymptotic in $\gamma_{RD}$ and $\gamma_{SD}$ following the approach in Eqn. (42) of~\cite{Jang2011}
\vskip-\parskip
\begin{align}
\label{eqn:P_1_lim}
P_1^{\gamma_{RD}} \triangleq & \lim_{\gamma_{RD}\to\infty} P_1 = \left( 1 - Q(\sqrt{2\gamma_{SR}})\right)Q\left[ \frac{\sqrt{2}(\gamma_{SD}+\gamma_{SR})}{\sqrt{\gamma_{SD}}}\right]\nonumber\\
P_1^{\gamma_{SR}} \triangleq & \lim_{\gamma_{SR}\to\infty} P_1 = Q\left[ \sqrt{2(\gamma_{SD}+\gamma_{RD})}\right]
\end{align}
to approximate $P_1$ with the sum of these two terms. In this way, $P_1$ is now the sum of two functions both of which have two arguments and are suitable for an approximation with impulse functions. It should be noted that in the approach utilized in Section~\ref{sec:Qfunc_min}, $Q(\sqrt{2x})$ and $Q(\sqrt{2y})$ are also asymptotic terms. Using the result of Section~\ref{sec:Qfunc_double}, taking $\sigma_{SD}^2 =\sigma_{SR}^2 = \sigma_{RD}^2=1$ and average SNR as $\bar{\gamma}$, approximate expectation of $P_1$ is evaluated to be
\vskip-\parskip
\begin{align}
\label{eqn:P_1}
I_3 \approx &  \int_{0}^{\infty} \int_{0}^{\infty} P_1^{\gamma_{RD}}f_{\gamma_{SR}}(\gamma_{SR})f_{\gamma_{SD}}(\gamma_{SD})\text{d}\gamma_{SR}\text{d}\gamma_{SD}
+ \int_{0}^{\infty} \int_{0}^{\infty} P_1^{\gamma_{SR}}f_{\gamma_{RD}}(\gamma_{RD})f_{\gamma_{SD}}(\gamma_{SD})\text{d}\gamma_{RD}\text{d}\gamma_{SD}\nonumber\\
\approx &\frac{1}{16\bar{\gamma}^2}\exp \left( -\frac{1.3049}{\bar{\gamma}}\right)+\frac{3}{16\bar{\gamma}^2}\exp \left( -\frac{2(0.8197)}{\bar{\gamma}}\right).
\end{align}

Defining similar asymptotic terms for $P_2$, we reach
\vskip-\parskip
\begin{align}
\label{eqn:P_2}
I_4 \approx &  \int_{0}^{\infty} \int_{0}^{\infty} P_2^{\gamma_{RD}}f_{\gamma_{SR}}(\gamma_{SR})f_{\gamma_{SD}}(\gamma_{SD})\text{d}\gamma_{SR}\text{d}\gamma_{SD}\nonumber\\
\approx &\frac{1}{4\bar{\gamma}^2}\exp \left( -\frac{1.7564+1.3737}{\bar{\gamma}}\right).
\end{align}

\begin{figure}[htbp]
   \centering
   \includegraphics[width=0.7\textwidth]{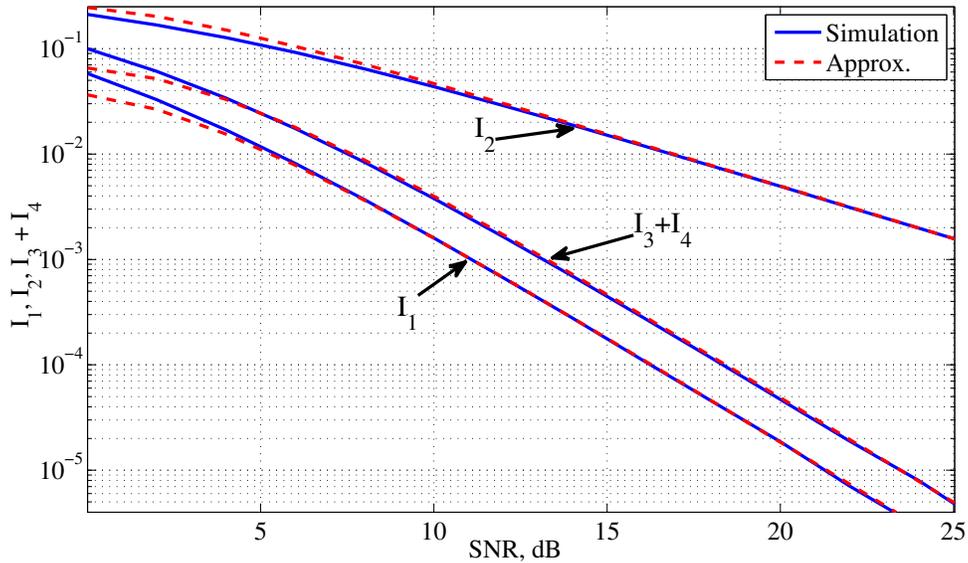}
   \caption{Approximating the integrals $I_1$, $I_2$, and $I_3+I_4$}
   \label{fig:Qcombined}
\end{figure}

Finally, summing the results of (\ref{eqn:P_1}) and (\ref{eqn:P_2}) we obtain and plot the approximate expectation of $P^b$ as $I_3+I_4$ in Fig.~\ref{fig:Qcombined} together with the simulation result. It is seen that the analysis proposed in this work yields an extremely good approximation to the end-to-end average BER of the canonical cooperative communication system by giving the closed form expression as a product of the coding and diversity gain terms. We initially observe that the $\exp \left( \frac{-c}{SNR}\right)$ terms will vanish as SNR tends to infinity, where the diversity order and the coding gain are defined. Hence, firstly, the diversity gain is taken as the exponent of the $\frac{1}{\bar{\gamma}}$ terms from (\ref{eqn:P_1}) and (\ref{eqn:P_2}) and is $2$ as found in~\cite{Wang07}. Then the coding gain is calculated as the sum of the constants multiplying the terms that are functions of SNR and is simply $1/2$.

\section{Sampling Property for Wireless Network Coding}
\label{sec:nwc}
The procedure followed in Section~\ref{sec:Qfunc_relay} for the analysis of canonical cooperative communication system is extended for the network coded system in this section. The analysis is also based on the receiver structure which utilizes the equivalent channel approximation for the two-hop links carrying the network coded bits. The difference from Section~\ref{sec:Qfunc_relay} is that now the receiver should also decode the network code for detecting $k$ source bits rather than detecting a single source's bit by applying simple MRC method on its weighted observation signals. We start with the network coded system description.

\subsection{Network Coded System Model}
\label{sec:nwc_system}
The network coded communication system investigated in this work is a sample scenario selected according to the model which is detailed in~\cite{Aktas2013}. In~\cite{Aktas2013}, source nodes transmit in orthogonal time slots and each source node serves potentially as a relay to the other source nodes. 
\begin{figure}[htbp]
   \centering
   \includegraphics[width=0.5\textwidth]{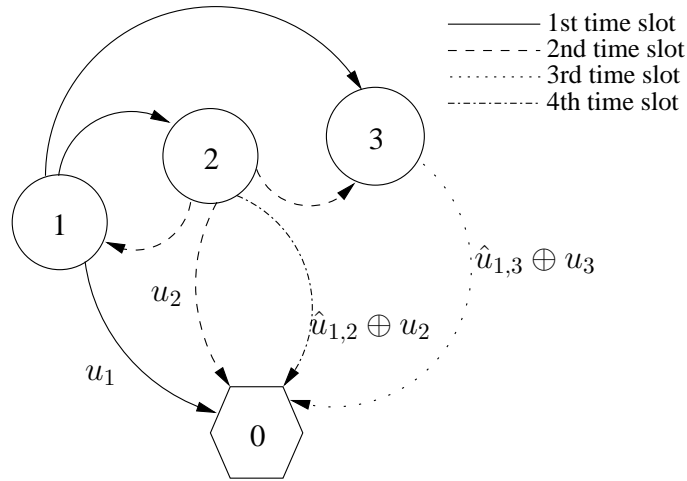}
   \caption{A sample network coded wireless communication scenario.}
   \label{fig:sample_nw}
\end{figure}
The specified network coding scenario is depicted in Fig.~\ref{fig:sample_nw}. In the sample network, $k=3$ source nodes are allowed to transmit to a separate destination, node $0$, following a time-division access method with a round of communication that comprises $n=4$ time slots. The source nodes $1$, $2$, and $3$ aim to transmit their data symbols $u_1$, $u_2$, and $u_3$ respectively. In this work, we assume $u_i$ to be binary for the sake of a simple system description and performance analysis, although all of the following arguments can be generalized for larger alphabets. The network coding rules followed are prescribed by the generator matrix $\mathbf{G}$ and the scheduling of the nodes to use the channel is given in the transmitting node vector $\mathbf{v}$:
\vskip-\parskip
\begin{align}
\label{eqn:gen}
\mathbf{G}=\left[ \begin{array}{cccc}
1 & 0 & 1 & 1 \\
0 & 1 & 0 & 1 \\
0 & 0 & 1 & 0 \end{array} \right], \mathbf{v} = \left[1\ 2 \ 3\ 2\right].
\end{align}

As an example, from Fig.~\ref{fig:sample_nw}, in the first time slot node $1$ transmits its own data bit $u_1$ using BPSK, which corresponds to the first entry of $\mathbf{v}$ and the first column of $\mathbf{G}$ respectively. In a similar fashion, in the second slot, node $2$ directly transmits own data bit to all other nodes in the network without any network coding. On the other hand, in the third slot, node $3$ combines its own bit $u_3$ with its estimate $\hat{u}_{1,3}$ obtained by using the DMF rule during the first time slot, where $u_1$ is transmitted. The combination is carried out as a simple XOR operation so that node $3$ modulates and transmits the network coded bit $\hat{u}_{1,3}\oplus u_3$. In the last slot, node $2$ transmits $\hat{u}_{1,2}\oplus u_2$ by using its own detection result $\hat{u}_{1,2}$ corresponding to $u_1$. In this work, we assume that all the channels associated to the links drawn in Fig.~\ref{fig:sample_nw} follow independent Rayleigh block fading (constant during a time slot) and the channel state information is only available to the receiving end of each link without any feedback in the network. In addition, as described in~\cite{Aktas2013}, the intermediate nodes help mitigate the problem of error propagation in the subsequent hops. As an example, the transmission of network coded bit $\hat{u}_{1,3}\oplus u_3$ is based on a possibly erroneous detection at the intermediate node $3$. Therefore the intermediate nodes are assumed to append the probability of error in their network coding operation to the transmitted packets so that the destination node $0$ can take the possible detection errors into consideration. Under these assumptions, node $0$, which is responsible for decoding the source nodes' data, collects the following $n=4$ observations
\vskip-\parskip
\begin{align}
\label{eqn:obs1}
y_1 =& h_1 \mu (u_1) + n_1,\nonumber\\
y_2 =& h_2 \mu (u_2) + n_2,\nonumber\\
y_3 =& h_3 \mu (u_1\oplus e_3\oplus u_3) + n_3,\nonumber\\
y_4 =& h_4 \mu (u_1\oplus e_4\oplus u_2) + n_4, 
\end{align}
where $h_j$, $j=1,\hdots,4$, denotes the independent channel gain coefficient for the $j$th slot and is assumed to follow $CN(0,1)$. Also $w_j$, $j=1,\hdots,4$, denotes the ZMCSCG noise signal term and is assumed to be independent from all other noise terms, channel coefficients, and data bits. Each noise term has a variance of $N_0$. We use the mapping $\mu(u_i)=\sqrt{E}(1-2u_i)$ with BPSK modulation. We further define the average SNR as $\bar{\gamma}=\frac{E}{N_0}$ and hence the instantaneous SNR corresponding to each time slot as $\gamma_{j}=\bar{\gamma} \vert h_{j}\vert ^2$, $j=1,\hdots,4$. Moreover, $e_j$ is the binary error term for the network coding operation in the $j$th slot. As a demonstration, if node $3$ has detected $u_1$ bit correctly in the first slot, we have $e_3=0$ and $e_3=1$ otherwise. Clearly, for the first two slots in which no network coding is utilized we should have $e_1=e_2=0$ deterministically. 

After receiving the signals given in~(\ref{eqn:obs1}), node $0$ tries to optimally decode the source data. As an example for the data bit of node $1$, the \textit{individual} Maximal A Posteriori (MAP) rule is given as
\vskip-\parskip
\begin{align}
\label{eqn:individual_det}
\hat{\hat{u}}_1 =& \argmax_{u_1}p(u_1\vert \mathbf{y})\nonumber\\
=& \argmax_{u_1}\sum_{u_2,\ldots ,u_k}\sum_{e_1,\ldots ,e_n} p(\mathbf{y} \vert \mathbf{u,e})\prod_{j=1}^{n}p(e_j),
\end{align}
where $\mathbf{y}=[y_1 \ y_2 \ y_3 \ y_4]^T$ is the observation vector, $\mathbf{u}=[u_1 \ u_2 \ u_3]$ is the source data vector and $\mathbf{e}=[0 \ 0 \ e_3 \ e_4]^T$ denotes the error term vector. The \textit{joint} MAP rule detecting the sequence $\mathbf{u}$ as a whole may also be preferable in terms of computational complexity and ease of analysis
\vskip-\parskip
\begin{align}
\label{eqn:joint_det}
\hat{\hat{\mathbf{u}}} = \argmax_{\mathbf{u}}\sum_{e_1,\ldots ,e_n} p(\mathbf{y} \vert \mathbf{u,e})\prod_{j=1}^{n}p(e_j).
\end{align}
Both of the rules in (\ref{eqn:individual_det}) and (\ref{eqn:joint_det}) are applicable in the equivalent channel approach that is utilized for the analysis of the end-to-end BER performance of the network coded system in Section~\ref{sec:equiv_analysis}. It is also important to note that a network decoder with linear-time complexity based on the sum-product algorithm was proposed in~\cite{Aktas2013}, which performs quite close to the optimal detection rule given in (\ref{eqn:individual_det}). 

\subsection{BER Analysis for a Network Coded System Using Equivalent Channel Approach}
\label{sec:equiv_analysis}
Let us start investigating (\ref{eqn:obs1}) with emphasis on the observations in the third and the fourth time slots in which network coded cooperation is utilized. For these observations, in order to simplify both the design of the receiver and the analysis, we define the equivalent instantaneous SNR values $\gamma_{eq3}=\bar{\gamma} \vert h_{eq3}\vert ^2$ and $\gamma_{eq4}=\bar{\gamma} \vert h_{eq4}\vert ^2$ as done in Section~\ref{sec:model}.
\vskip-\parskip
\begin{align}
\label{eqn:equiv}
\vert \gamma_{eq3}  \vert ^2=\frac{Q^{-1}\left\lbrace p_{e3}\left(1-Q\left(\sqrt{2\vert \gamma_{3}\vert ^2}\right)\right) + \left(1-p_{e3}\right)Q\left(\sqrt{2\vert \gamma_{3} \vert ^2}\right)\right\rbrace}{2},
\end{align}
where $p_{e3}$ denotes the probability that node $3$ detects $u_1$ in error (hence forwards an erroneous network coded bit) and with BPSK modulation it is  found as $p_{e3}=Q\left(\sqrt{2\vert \gamma_{1\rightarrow 3}\vert ^2}\right)$ with given SNR value $\gamma_{1\rightarrow 3}$ for the link from node $1$ to $3$ in the first time slot. Similarly, another equivalent gain $\gamma_{eq4}$ is defined and the weighted observation signal vector at the receiver is found as
\vskip-\parskip
\begin{align}
\label{eqn:obs2}
z_1 =& w_1 y_1 = h_1^\ast y_1 = \vert h_1 \vert ^2 \mu (u_1) + h_1^\ast n_1,\nonumber\\
z_2 =& w_2 y_2 = h_2^\ast y_2,\nonumber\\
z_3 =& w_3 y_3 = \frac{\gamma_{eq3}  }{\gamma_{3} }h_3^\ast y_3,\nonumber\\
z_4 =& w_4 y_4 = \frac{\gamma_{eq4}  }{\gamma_{4} }h_4^\ast y_4.
\end{align}
In (\ref{eqn:obs2}), the weights for the first two observations are exactly the same as those of the MRC technique, since both of these are direct transmissions with no network coding. Given the equivalent channel outputs in (\ref{eqn:obs2}), one can implement both the joint and the individual detection rule given in Section~\ref{sec:nwc_system} for detection of data vector $\mathbf{u}$. The individual detection rule for $u_i$ is
\vskip-\parskip
\begin{align}
\label{eqn:indiv}
\hat{\hat{u}}_1=\argmax_{u_1\in \{ 0,1\} }\sum_{u_2,u_3\in \{ 0,1\} }p(\mathbf{z}\ \vert\ \mathbf{u}),
\end{align}
while the joint detection rule for $\mathbf{u}$ is found as
\vskip-\parskip
\begin{align}
\label{eqn:joint}
\mathbf{\hat{\hat{u}}}=\argmax_{\mathbf{u}\in \{ 0,1\}^k }p(\mathbf{z}\ \vert\ \mathbf{u}).
\end{align}
Although both of these detectors are very close in terms of BER performance to the optimal detector of (\ref{eqn:individual_det}) as to be shown in Section~\ref{sec:sim_results}, the joint detector in (\ref{eqn:joint}) is much simpler to analyze (and also to implement) due to absence of the summation operation over all possible data vectors. As a consequence, we aim to reach an expected BER expression for the joint detector using the sampling property of the Q-function. This expression is going to be an upper bound for the optimal detector in (\ref{eqn:individual_det}).

We start the analysis of the detector in (\ref{eqn:joint}) by identifying the conditional probability distributions (for given respective channel coefficients) of the weighted observations. Without loss of generality, we take $N_0=1$ for the remaining part of the paper in order to simplify the derivation. The transmissions in the first two time slots are direct, hence the distributions of weighted observations are relatively simple. As an example, by further conditioning the observations on the input data vector pattern $u_1=u_2=u_3=0$, one finds
\vskip-\parskip
\begin{align}
\label{eqn:pdf_exact_direct}
(z_j \vert u_1=u_2=u_3=0) \sim CN \left(\gamma_{j},\gamma_{j} \right), \; j\in\left\lbrace 1,2 \right\rbrace,
\end{align}
where the Gaussian distribution of the noise signal is used. These distributions are also used in the detector that makes use of the equivalent channel approach. On the other hand, for the slots making use of network coding, we obtain the following conditional distribution for the given data and error pattern $u_1=u_2=u_3=e_3=e_4=0$:
\vskip-\parskip
\begin{align}
\label{eqn:pdf_exact_nc}
(z_j \vert u_1=u_2=u_3=e_j=0) \sim CN \left(\gamma_{eq\,j} ,\frac{\gamma_{eq\,j}^2}{\gamma_{j}}\right),  \; j\in\left\lbrace 3,4\right\rbrace.
\end{align}
However, the equivalent channel detector uses the following distributions (without any consideration on the relay error variables $e_j$) in the construction of the detection rules
\vskip-\parskip
\begin{align}	
\label{eqn:equiv_dist}
(z_j & \vert u_1=u_2=u_3=0) \sim CN \left(\gamma_{eq \, j},\gamma_{eq \, j} \right), \; j\in\left\lbrace 3,4\right\rbrace.
\end{align}
since it assumes the network coded data signals are being transmitted over the equivalent channel with no relay error but decreased instantaneous SNR, $\gamma_{eq \, j}$. As a result, in the analysis of the conditional BER for $u_1=u_2=u_3=e_3=e_4=0$, we derive the detection rules according to (\ref{eqn:equiv_dist}), whereas the probability of an error for a detection rule is calculated using (\ref{eqn:pdf_exact_nc}) for the time slots $3$ and $4$. Similar distributions can also be found for the condition $e_j=1$. 

Due to linearity of the block code used for constructing the network code in the system, one can assume all-zero data vector transmission, $u_1=u_2=u_3=0$, and find the probability of error for a given data symbol. As an example, for $u_1$ we have the following upper bound for end-to-end bit error probability: 
\vskip-\parskip
\begin{align}
\label{eqn:union}
P(\hat{\hat{u}}_1 \neq u_1) &= \sum_{u_2,u_3}P(\mathbf{\hat{\hat{u}}}=[1\ u_2\ u_3]\ \vert \ \mathbf{u}=[0\ 0\ 0])\nonumber\\
&\leq \sum_{u_2,u_3} P(p_{eq}(\mathbf{z}\ \vert \ [0\ 0\ 0])< p_{eq}(\mathbf{z}\ \vert \ [1\ u_2\ u_3])),
\end{align}
where we make use of the union bound in the inequality. In (\ref{eqn:union}) $p_{eq}$ denotes the pdf utilized by the detector and is found for any data vector $\mathbf{u}$ in a similar fashion to (\ref{eqn:equiv_dist}):
\vskip-\parskip
\begin{align}
\label{eqn:pdf_eq}
p_{eq}(\mathbf{z}\ \vert \ \mathbf{u}))=\prod_{j=1}^4 {\frac{\exp\left(-\frac{\left\vert z_j - \left(-1\right)^{\mathbf{u} \mathbf{g}_j} \gamma_{eq \, j}\right\vert ^2}{\gamma_{eq \, j}}\right)}{\pi \gamma_{eq \, j}}},
\end{align}
where $\mathbf{g}_j$ is the $j^{\text{th}}$ column of the network code generator matrix $\mathbf{G}$ and $\gamma_{eq \, j}=\gamma_j$ for direct transmissions, $j=1,2$. We then condition $P(p_{eq}(\mathbf{z}\ \vert \ [0\ 0\ 0])< p_{eq}(\mathbf{z}\ \vert \ [1\ u_2\ u_3]))$ on the intermediate node error vectors. For the condition $e_3=e_4=0$, by using (\ref{eqn:pdf_eq}), we obtain 
\vskip-\parskip
\begin{align}
\label{eqn:cond_pdf_eq}
P \left( p_{eq}(\mathbf{z}\ \vert \ [0\ 0\ 0])< p_{eq}(\mathbf{z}\ \vert \ [1\ u_2\ u_3]) \big\vert e_3=e_4=0 \right) =&
P\left(\prod_{j=1}^4 \exp\left(-\frac{\vert z_j - \gamma_{eq \, j}\vert ^2}{\gamma_{eq \, j}}\right) \right. < \nonumber\\
&\phantom{P\left(\right.}\left.\prod_{j=1}^4 \exp\left(-\frac{\vert z_j - \left(-1\right)^{\left([1\, u_2\, u_3]g_j\right)} \gamma_{eq \, j}\vert ^2}{\gamma_{eq \, j}}\right)\right)
\end{align}
As an example, for the erroneously detected vector $\mathbf{\hat{\hat{u}}}=[1\ 0\ 0]$, we rewrite (\ref{eqn:cond_pdf_eq}) as
\vskip-\parskip
\begin{align}
P\left( p_{eq}(\mathbf{z}\ \vert \ [0\ 0\ 0])< p_{eq}(\mathbf{z}\ \vert \ [1\ 0\ 0]) \big\vert e_3=e_4=0 \right) =&
P\left(4\left[ RE\left\lbrace z_1 + z_3 + z_4 \ \right\rbrace \right]< 0\right)\nonumber\\ 
=& Q\left( \frac{\sqrt{2}\left( \gamma_{1} + \gamma_{eq3} + \gamma_{eq4} \right)}{\sqrt{\gamma_{1} + \frac {\gamma_{eq3}^2}{\gamma_{3}} + \frac {\gamma_{eq4}^2}{\gamma_{4}}}} \right),
\end{align}
where we used (\ref{eqn:pdf_exact_direct}) and (\ref{eqn:pdf_exact_nc}) in the last identity. In the next step of derivation for instantaneous BER expression, we sum up the conditional probability terms for all data vectors in error to reach the conditional version of (\ref{eqn:union}) with $e_3=e_4=0$ as
\vskip-\parskip 
\begin{align}
\label{eqn:u1_hat_neq_u1}
P(\hat{\hat{u}}_1 \neq u_1 \big\vert e_3=e_4=0 ) \leq & Q\left( \frac{\sqrt{2}\left(\gamma_{1} + \gamma_{eq3} + \gamma_{eq4} \right)}{\sqrt{\gamma_{1} + \frac {\gamma_{eq3}^2}{\gamma_{3}} + \frac {\gamma_{eq4}^2}{\gamma_{4}}}} \right)
+ Q\left( \frac{\sqrt{2}\left(\gamma_{1} + \gamma_{eq4} \right)}{\sqrt{\gamma_{1} + \frac {\gamma_{eq4}^2}{\gamma_{4}}}} \right) \nonumber\\
+ & Q\left( \frac{\sqrt{2}\left(\gamma_{1} + \gamma_{2} + \gamma_{eq3} \right)}{\sqrt{\gamma_{1} + \gamma_{2} + \frac {\gamma_{eq3}^2}{\gamma_{3}}}} \right)
+ Q\left(\sqrt{2\left(\gamma_{1} + \gamma_{2}\right)} \right).
\end{align}
Finally, by obtaining and weighting the conditional error probabilities for all error vector patterns we get
\vskip-\parskip
\begin{align}
\label{eqn:inst_ber}
 P  (\hat{\hat{u}}_1 \neq u_1) \simeq (1-p_{e4}) Q\left( \frac{\sqrt{2}\left(\gamma_{1} + \gamma_{eq4} \right)}{\sqrt{\gamma_{1} + \frac {\gamma_{eq4}^2}{\gamma_{4}}}} \right) 
 + p_{e4} Q\left( \frac{\sqrt{2}\left(\gamma_{1} - \gamma_{eq4} \right)}{\sqrt{\gamma_{1} + \frac {\gamma_{eq4}^2}{\gamma_{4}}}} \right)
 + Q\left(\sqrt{2\left(\gamma_{1} + \gamma_{2} \right)} \right),
\end{align}
where the terms which decrease with $\bar{\gamma}^3$ are neglected. If we use the two-dimensional sampling property results of Section~\ref{sec:Qfunc} in finding the expectation of the instantaneous BER (\ref{eqn:inst_ber}), we approximate the average BER expression for $u_1$ with
\vskip-\parskip
\begin{align}
\label{eqn:avg_ber}
E_{h_1,h_2,h_4,p_{e4}}\left\lbrace  P(\hat{\hat{u}}_1 \neq u_1)\right\rbrace \approx \frac{1}{16\bar{\gamma}^2}\exp \left( -\frac{1.3049}{\bar{\gamma}}\right)
 + \frac{3}{8\bar{\gamma}^2}\exp \left( -\frac{1.6394}{\bar{\gamma}}\right)
 + \frac{4}{16\bar{\gamma}^2}\exp \left( -\frac{3.1301}{\bar{\gamma}}\right).
\end{align}
Following the same procedure one can obtain the end-to-end BER expressions for $u_2$ ve $u_3$ as well. Agreement of these derived expressions with the simulation results is shown in Section~\ref{sec:sim_results}.

\section{SIMULATION RESULTS}
\label{sec:sim_results}
\begin{figure}[htbp]
   \centering
   \includegraphics[width=0.7\textwidth]{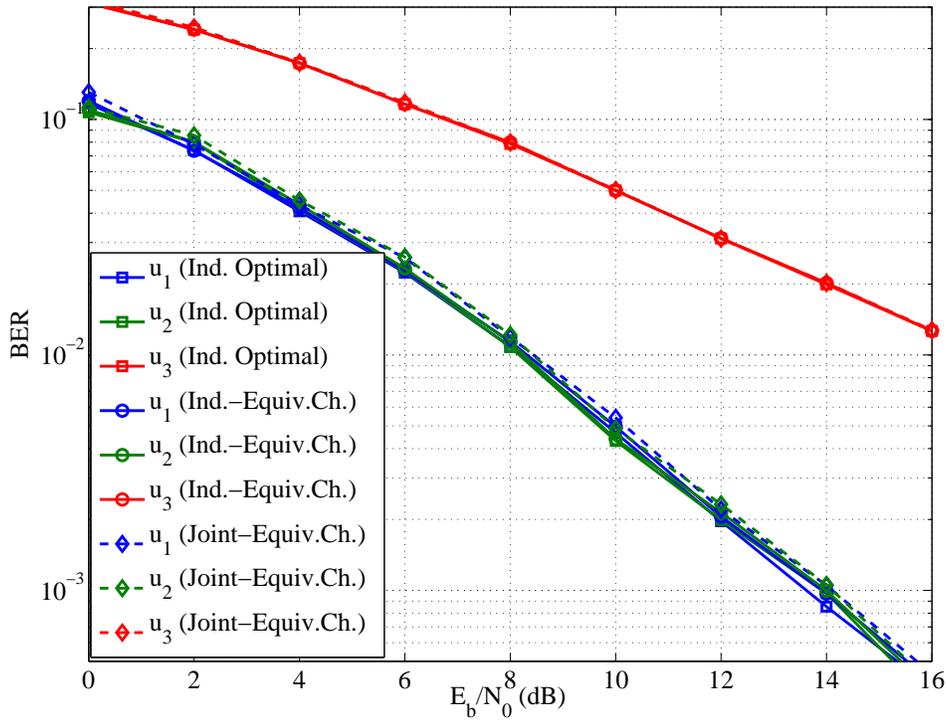}
   \caption{Performance of the optimal and the equivalent channel decoders}
   \label{fig:opt_vs_equiv}
\end{figure}

We present the performance figures in this section in order to validate the analysis for the network coded cooperative communication system done in Section~\ref{sec:nwc}. We give the results for the network coded system introduced in Section~\ref{sec:nwc_system} with Fig.~\ref{fig:sample_nw}. Initially, we observe the comparison of BER curves for three network decoders based on simulations in Fig.~\ref{fig:opt_vs_equiv}: the optimal decoder of (\ref{eqn:individual_det}), the suboptimal equivalent channel individual decoder of (\ref{eqn:indiv}), and the suboptimal equivalent channel joint decoder of (\ref{eqn:joint}). Based on this observation, we state that both the equivalent channel assumption and the joint detection simplification have negligible effect on the BER performance of network decoding operation. Thus a valid analysis for the simplest decoder of (\ref{eqn:joint}) serves as a good performance metric for the optimal decoder of (\ref{eqn:individual_det}) as well.

\begin{figure}[htbp]
   \centering
   \includegraphics[width=0.68\textwidth]{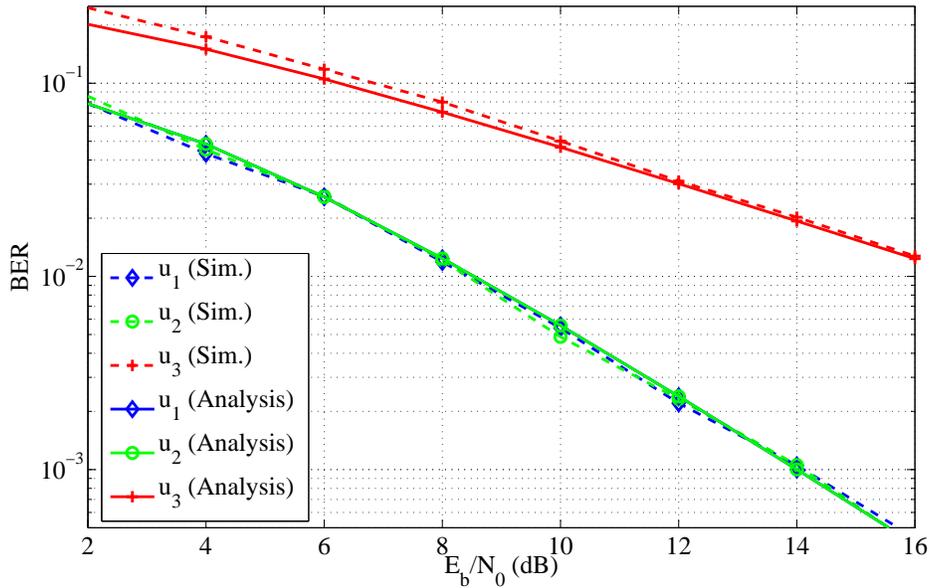}
   \caption{Simulation and analysis results for equivalent channel joint decoder}
   \label{fig:equiv_analysis}
\end{figure}

In Fig.~\ref{fig:equiv_analysis}, we present the agreement between the BER curves of the simulations for the joint equivalent channel network decoder of (\ref{eqn:joint}) and the BER expressions we derive in Section~\ref{sec:equiv_analysis}. The simulated performance curves for all data bits are in good agreement with the analysis results for a wide range of SNR values. As a result, the analysis that is using the generalized forms of the sampling property for the Q-function perfectly fits the simulation results for this network coded system of interest. 

\section{CONCLUSION and DISCUSSION}
\label{sec:concl}
Some useful average BER expressions for both the basic relayed and the network coded communication scenarios under Rayleigh block fading are derived in closed form. In order to obtain these closed form expressions, we generalize the sampling property of the Q-function that is frequently observed in instantaneous BER functions of these systems. The generalization includes an insightful analysis of the applicability of the sampling function in addition to the extension of it to integrands of more than one variables. We also propose a network decoder which operates under equivalent channel assumption. By combining two independent paths from a source and an intermediate node to the destination under a single channel, this assumption enables both reduction in the complexity of the analysis and the decoding with a negligible loss in performance. We substantiate the validity of these generalizations on sampling property for integrals and the use of equivalent channel assumption in network coded systems through extensive simulations. The adaptation of these methods to other types of fading channels and to scenarios including correlated channels seem to be interesting open problems.

\bibliography{Tugcan_Aktas_WNC_Analysis_TCOM_onecolumn}

\end{document}